\documentclass[aps,pre,twocolumn,reprint,amsmath,amssymb,superscriptaddress,showpacs,floatfix]{revtex4-1}
\usepackage{graphicx}
\usepackage{dcolumn}
\usepackage{bm}
\usepackage[colorlinks, allcolors=blue]{hyperref}
\usepackage[usenames, dvipsnames]{color}

\begin{document}

\title{Quantum phase diagram of a frustrated spin-1/2 system on a Trellis Ladder}

\author{Debasmita Maiti}
\email{debasmita.maiti@bose.res.in}
\affiliation{S. N. Bose National Centre for Basic Sciences, Block - JD, Sector - III, Salt Lake, Kolkata - 700106, India}

\author{Manoranjan Kumar}
\email{manoranjan.kumar@bose.res.in}
\affiliation{S. N. Bose National Centre for Basic Sciences, Block - JD, Sector - III, Salt Lake, Kolkata - 700106, India}

\date{\today}

\begin{abstract}
        We study an isotropic Heisenberg spin-1/2 model on a trellis ladder which is composed of two $J_1-J_2$ zigzag ladders 
	interacting through anti-ferromagnetic rung coupling $J_3$. The $J_1$ and $J_2$ are ferromagnetic zigzag spin 
	interaction between two legs and antiferromagnetic interaction along each leg of a zigzag ladder. A quantum phase 
	diagram of this model is constructed using the density matrix renormalization group (DMRG) method and linearized 
	spin wave analysis. In small $J_2$ limit a short range striped collinear phase is found in the presence of $J_3$, whereas, 
	in the large $J_2/J_3$ limit non-collinear quasi-long range phase is found. The system shows a short 
	range non-collinear state in large $J_3$ limit. The short range order phase is the dominant feature of this phase diagram.
        We also show that the results obtained by DMRG and linearized spin wave analysis show similar phase boundary 
        between collinear striped and non-collinear short range phases, and the collinear phase region shrinks with increasing 
	$J_3$. We apply this model to understand the magnetic properties of CaV$_2$O$_5$ and also fit the experimental data of susceptibility
        and magnetization. 
        We note that $J_3$ is a dominant interaction in this material, whereas $J_1$ and $J_2$ are approximately half of $J_3$.
        The variation of magnetic specific heat capacity as a function of temperature for various external magnetic fields 
        is also predicted.           
\end{abstract}

\maketitle

\section{\label{sec:intro}Introduction}
In the last couple of decades frustrated low dimensional quantum magnets
have been intensively explored in search of various exotic phases
like spin fluid with quasi-long-range order (QLRO) ~\cite{OKAMOTO1992,Haldane1982,white96,chitra95,mk2015,mk2016},
spin dimer with short-range order (SRO) ~\cite{ckm69a,Shastry1981,Haldane1982,white1994,white96},
vector chiral ~\cite{chubukov1991,furukawa2012}, multipolar
phases ~\cite{Zhitomirsky2010,chubukov1991,furukawa2012,Aslam2017} etc. These phases
arise in presence of some specific types of spin exchange interactions which may enhance the quantum
fluctuations in low-dimensional frustrated systems like
one dimensional (1D) spin chains realized in materials,
LiCuVO$_4$ ~\cite{Mourigal2012}, Li$_2$CuZrO$_4$ ~\cite{Drechsler2007},
Li$_2$CuSbO$_4$ ~\cite{dutton_prl}, (N$_2$H$_5$)CuCl$_3$~\cite{Maeshima2003} etc., and quasi-1D
spin ladders manifested in form of SrCu$_2$O$_3$~\cite{sandvik1995},
(VO)$_2$P$_2$O$_7$~\cite{dcjhon1987,dagotto96} etc. Frustrated twisted ladders are also realized in materials like
Ba$_3$Cu$_3$In$_4$O$_{12}$ and Ba$_3$Cu$_3$Sc$_4$O$_{12}$ ~\cite{Dutton2012,Kumar2013,Volkova2012,Koteswararao2012}.
Majority of the 1D frustrated magnetic systems mentioned above are modelled by simple $J_1$-$J_2$ chain
~\cite{ckm69a,ckm69b,Shastry1981,Tonegawa1987,Kuboki1987,Hamada1988,OKAMOTO1992,chitra95,Bursill1995,white96,Allen1997,Itoi2001,
Hase2004,Lu2006,Drechsler2007,mk2015,mk2016}. 
This model can explain the gapless spin fluid ~\cite{OKAMOTO1992,chitra95}, gapped dimer ~\cite{ckm69a,white96},
gapped non-collinear ~\cite{chitra95,white96,mk2015,mk2016} and decoupled phases ~\cite{mk2010a}.

In fact many of these 1D systems like LiCuVO$_4$~\cite{Mourigal2012}, Li$_2$CuZrO$_4$~\cite{Drechsler2007}
show three dimensional ordering at low temperature;
therefore, interchain
couplings are considered to understand the interesting physics below the three
dimensional ordering temperature.
However, there are materials with effective spin interactions confined to quasi-1D ladder like structure e.g.,
SrCu$_2$O$_3$~\cite{sandvik1995}, (VO)$_2$P$_2$O$_7$ ~\cite{dcjhon1987,dagotto96},
CaV$_2$O$_5$, MgV$_2$O$_5$ ~\cite{Korotin1999,Tdasgupta2000} etc. These systems have 
antiferromagnetic (AFM) spin exchange interactions along both legs and rungs, and 
there is also a weak interaction between two adjacent ladders. 
The ground state (gs) of these systems is a gapped SRO phase ~\cite{white1994}. 
The coexisting of spin gap and long range magnetic order in the ladder compound 
LaCuO$_{2.5}$ is explained considering interladder coupling ~\cite{Normand1996,Troyer1997}.

\begin{figure}
\centering
\includegraphics[width=3.0 in]{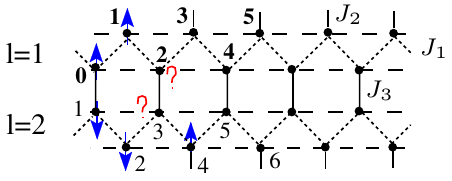}
\caption{\label{fig:figure1} Two coupled zigzag ladders form trellis
ladder. The extended lines show the extension of trellis ladder to a 2D
trellis lattice structure. The arrows represent arrangement of spins and question marks represent frustrated spins.
The reference site is labeled by '\textbf{0}' and the distances of other sites along
same ladder are shown in bold numbers, and normal numbers represent the distances on the other ladder.
$l$ represents the zigzag ladder indices.}
\end{figure}

The 1D $J_1-J_2$ system, in large $J_2$ limit, is called zigzag ladder~\cite{chitra95}, where
two chains are coupled through zigzag bonds, for example LiCuV$O_4$ ~\cite{Mourigal2012}. The isolated ladders like
zigzag and normal ladders have been extensively studied~\cite{white96,white1994,almeida2007,vekua2003,Agrapidis2017}; 
however, the effect of interladder coupling on these ladders is rarely studied. 
Networks of the coupled zigzag ladders can form a trellis lattice like structure as shown in Fig.~\ref{fig:figure1}.
The trellis lattice is composed of a number of zigzag ladders coupled through normal rung bonds;
alternatively, we can assume coupled normal ladders interacting through zigzag like bond interactions.
In this lattice spin exchange interaction strengths $J_2$ and $J_3$ are along leg and rung 
of a normal ladder, respectively, and $J_1$ is zigzag bond interaction strength between two ladders as shown in Fig.~\ref{fig:figure1}.

In this paper we consider only two coupled zigzag ladders and call it {\it trellis ladder}
because of its geometry. We also impose periodic boundary condition along the width to mimic the trellis lattice.
In various interaction limits, two coupled zigzag ladders can behave like
a two-leg honeycomb ladder as considered in ref.~\cite{Luo2018}, where both $J_1$ and $J_2$ are AFM,
but $J_3$ can be either ferromagnetic (FM) or AFM. This system shows two types of Haldane phase
for the FM $J_3$, and columnar dimer and rung singlet phases in presence of the AFM $J_3$.
Normand {\it et al.} have considered a similar coupled ladders with all three AFM
$J_1$, $J_2$ and $J_3$ interactions~\cite{Normand1997}. For large $J_2/J_1$ they have noticed dimerized gs, whereas
 non-collinear (NC) long-range order (LRO) for large $J_3$ ($J^\prime_2$). 
They have found N\'eel LRO phase in the small $J_2 < 0.4$ limit.
Zinke {\it et al.} have shown the effect of interchain coupling
on NC gs of $J_1-J_2$ model~\cite{Zinke2009}, in a two dimensional geometry. The effect of
interladder coupling on spin gap and magnon dispersion is calculated 
using perturbation theory by Miyahara {\it et al.}~\cite{Miyahara1998}.
They also try to model the magnetic susceptibility of SrCu$_2$O$_3$ and CaV$_2$O$_5$ using
quantum Monte Carlo and mean field type scaling methods~\cite{Miyahara1998}. However, 
the system with FM $J_1$, and AFM $J_2$ and $J_3$ has not been studied in the ladder geometry.
In this paper, we consider a spin-1/2 trellis ladder structure, which is composed of two zigzag ladders with FM $J_1$ and AFM $J_2$,
and they are coupled by AFM $J_3$ as shown in Fig.~\ref{fig:figure1}.
Our main focus of this paper is to construct the quantum phase diagram (QPD) and also 
understand the effect of rung interaction $J_3$ on the various exotic
phases of zigzag ladder~\cite{mk2015}. We notice that in small $J_2/|J_1|$ limit, gs
has collinear striped (CS) SRO on each zigzag ladder; however,
spins on one zigzag ladder are aligned antiferromagnetically with respect to the spins on the other zigzag ladder~\cite{MAITI2019}.
The NC spin order sets in for moderate value of $J_2$. 
The presence of QLRO
in NC regime at small $J_3/J_2$ limit is a striking effect of the $J_3$. In large $J_3$ limit, rung dimer
is the dominant gs.

This paper is divided into four sections. In section~\ref{sec2} the model Hamiltonian
and numerical method are explained. The numerical results are given in section~\ref{sec3}.
Linear spin wave analysis and experimental data fitting of Ca$V_2$$O_5$ are given in section~\ref{sec4} and section~\ref{sec5}, respectively.
All the results are discussed and summarized in Section~\ref{sec6}.

\section{\label{sec2}Model Hamiltonian and Numerical Method}
A four-legged spin-1/2 ladder made of two coupled zigzag ladders is considered as shown in Fig.~\ref{fig:figure1}.
The exchange interactions between spins along the legs and rungs are AFM in nature. The diagonal exchange interactions
$J_1$ in a zigzag ladder are FM. We can write an isotropic Heisenberg spin-1/2 model
Hamiltonian for the trellis ladder system as

\begin{eqnarray}
\label{eq:ham}
\mathrm{H} &= &\sum_{l = 1,2}\sum_{i=1}^{N/2} \ J_1 \, \vec{S}_{l,i} \cdot \vec{S}_{l,i+1} 
+ J_2 \, \vec{S}_{l,i} \cdot \vec{S}_{l,i+2} \nonumber \\
& & \qquad \qquad + J_3 \, \vec{S}_{1,i} \cdot \vec{S}_{2,i} + HS^z_i,
\end{eqnarray}
where $l=1,2$ are the zigzag ladder indices. $\vec{S}_{l,i}$ is the spin operator at site
$i$ on zigzag ladder $l$. We consider $J_1=-1$, and $J_2$ and $J_3$ are
variable AFM exchange interaction strengths. We use periodic boundary condition along the rungs,
whereas it is open along the legs of the system. 

We use density matrix renormalization group (DMRG) method to handle the large degrees of freedom in our system. This method is
a state of art numerical technique for 1D or quasi-1D system, and it is based on the systematic
truncation of irrelevant degrees of freedom ~\cite{white-prl92,karen2006,schollwock2005}.
We use recently developed DMRG method where four new sites are added at every
DMRG step~\cite{mk2010b}. This method while constructing superblock,
avoids the old-old operator multiplication which leads to the generation of large number of non-zero but small
matrix elements in superblock Hamiltonian. The number of eigenvectors $m$ corresponding to the largest eigenvalues
of the density matrix, is kept for the renormalization of operators and Hamiltonian of the system block.
We have kept $m$ up to 400 to restrict the truncation error less than $10^{-10}$. We have
used system sizes up to $N=300$ to minimize the finite size effect.

\section{\label{sec3}Results}
We first present an outline of the QPD which is constructed based on various
quantities like correlation function $C(r)$, pitch angle $\theta$ and bond order $C(r=1)$. The detailed
numerical and analytical calculations are discussed in the following subsections.
For $J_1=0$, this system is composed of two isolated normal ladders, and two isolated
zigzag ladders for $J_3=0$. In $J_1=0$ limit, gs shows the formation of singlet dimers along the rungs on the normal 
ladder~\cite{white1994}. On the other hand, for $J_3=0$ the system
shows various phases arising due to the presence of frustration in each zigzag ladder,
at different exchange coupling limits. For $J_2/|J_1| < 0.25$, the gs of an isolated zigzag ladder has ferromagnetically ordered spins
and gapless excitations. In the intermediate parameter regime,
$0.25 < J_2/|J_1| < 0.67$, NC order arises in this system with a small finite spin gap~\cite{chitra95,mk2016,mk2015,mk2013}.
The system behaves like decoupled AFM chains exhibiting QLRO in spin-spin correlation and gapless excitations in $J_2/|J_1|>0.67$
limit ~\cite{mk2016}. We notice that if two zigzag ladders start interacting with
each other through rung coupling $J_3$, it immediately opens a spin gap in the system. The spin gap in CS(SRO) phase has been explicitly
studied in ref.~\cite{MAITI2019}.
In section~\ref{sec4} we discuss the linear spin wave analysis of this model. At the end, we apply this model to fit
magnetic susceptibility and magnetization of Ca$V_2$$O_5$ in large $J_3$ limit. We also predict the specific heat curve at high temperature
which can be verified experimentally.
\begin{figure}
\centering
\includegraphics[width=4.0 in]{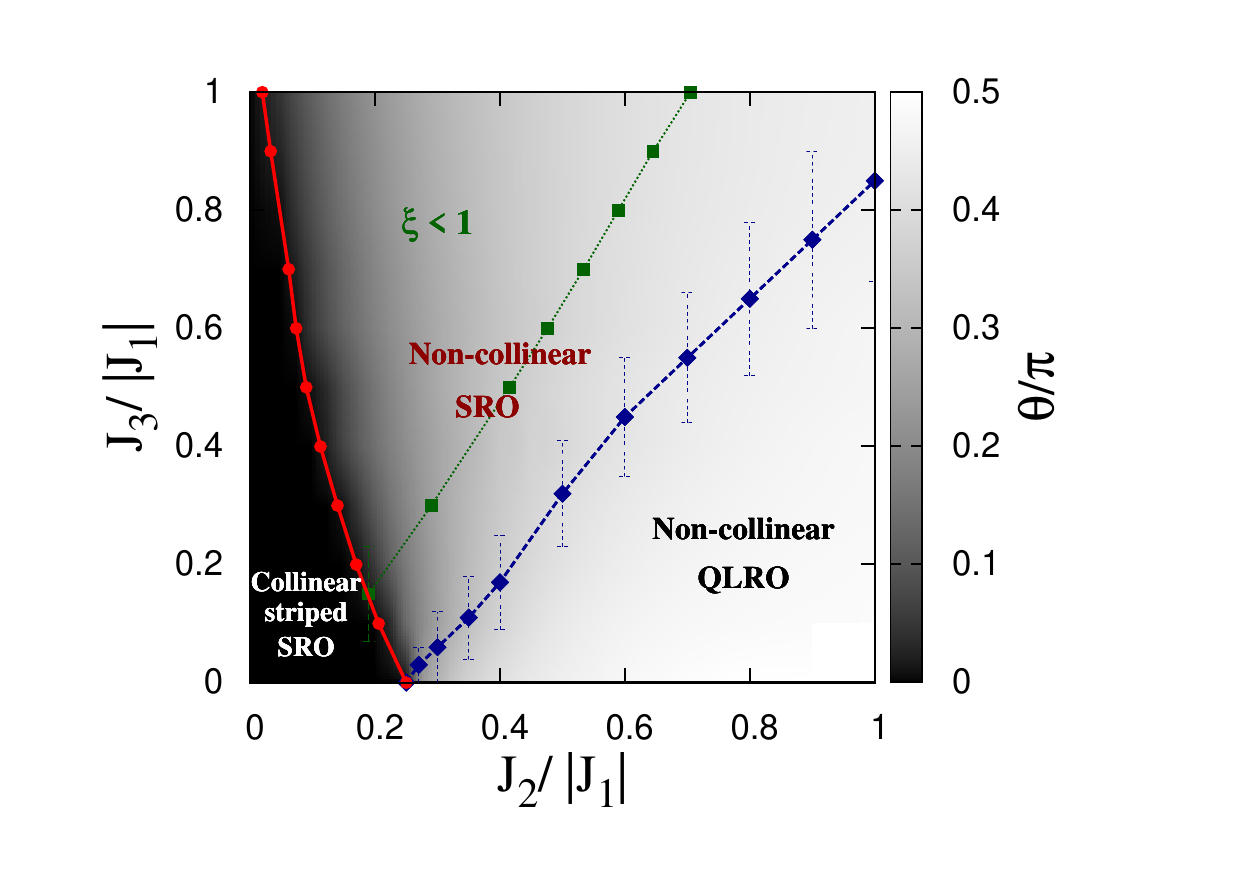}
\caption{\label{fig:phase}The QPD of the model Hamiltonian in 
Eq.~\ref{eq:ham} for $H = 0$:  red solid line with circles represents the boundary between CS(SRO) and NC(SRO) phases.
The green dotted line with square symbols in NC(SRO) regime represents the boundary line with $\xi \approx 1$.
NC(QLRO) phase lies below the blue dashed line with diamonds. The color gradient represents the
pitch angle $\theta$ distribution in the $J_2-J_3$ parameter space.
}
\end{figure}

\subsection{\label{sec21}Quantum Phase Diagram}
The quantum phase diagram of Hamiltonian in Eq.~\ref{eq:ham} is shown in $J_2/|J_1|$ and
$J_3/|J_1|$ parameter space, and we focus mainly on the phases in presence of $J_3$. 
The resulting phase diagram in Fig.~\ref{fig:phase} shows two
distinct phases: the CS(SRO) and NC spin order. In small $J_3$ and $J_2 < 0.25$ limit,
individual zigzag ladder retains the FM arrangement of spins;
however, the spins on two different zigzag ladders are aligned
antiparallelly with respect to each other. Therefore, the gs of the 
whole system has effective multiplicity $S^z=0$.
The spin-spin correlation decays exponentially along each zigzag chain. This phase can be called as CS
(SRO) phase.
As we increase $J_3$, the correlation length $\xi$ decreases. The details of this phase have been discussed already in the 
ref.~\cite{MAITI2019}. At higher $J_3$ value, even for $J_2 < 0.25$,
NC phase emerges but with small amplitude and $\xi$ in spin spin correlation. For $J_2 > 0.25$,
spiral arrangement of spins becomes more prominent for lower $J_3$. In the NC regime, C(r)
is either QLRO (decay following power law) called as NC(QLRO),
or SRO (exponentially decaying) called as NC(SRO)
for the small or large $J_3$, respectively. The $\theta$ vanishes at the boundary
between CS(SRO) and NC(SRO) phases. In Fig.~\ref{fig:phase}, color gradient represents $\theta$ distribution in the parameter
space. The red solid line with circles represents the boundary between CS(SRO) and NC(SRO) phases in the gs.
The region above the green dotted line with square symbols represents the SRO phase where spin correlation length is confined to
its neighbor i.e., $\xi \le 1$. In the large $J_3$ limit, the correlation strength along rung dominates,
and it tends to form singlet dimers along the rungs.
The dimer phase is characterized by large energy gap, and the spin correlation
is confined within the nearest neighbors ($\xi\le 1$).
Interestingly, for large $J_{2}/J_{3}$ limit the gs is in unique NC(QLRO) phase.
To best of our knowledge, QLRO phase exits with pitch angle $\theta =\pi$ or $\frac{\pi}{2}$ ~\cite{mk2015,mk2016}, whereas this system shows
QLRO even with $\theta<\frac{\pi}{2}$. NC(QLRO) phase lies below the blue dashed line with diamond symbols. The phase boundary between
NC(SRO) and NC(QLRO) phases has large errorbar due to the inability to distinguish between the power law and exponential nature of
$C(r)$ in this parameter regime. 
To verify these different phases $C(r)$, $\theta$, $\xi$ and $C(r=1)$ are studied in detail in the following subsections.

\begin{figure}
\centering
\includegraphics[width=3.0 in]{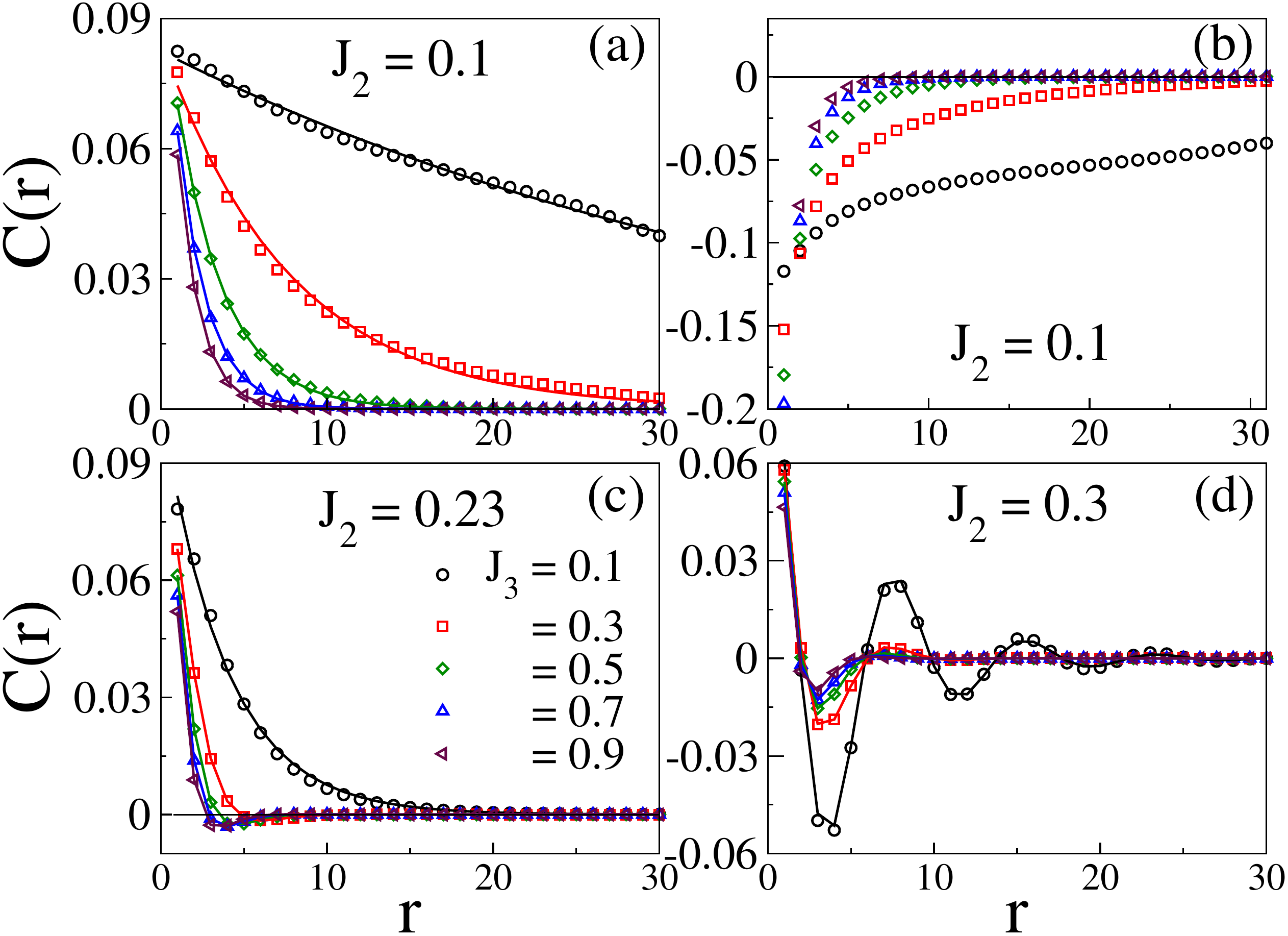}
\caption{\label{fig:cor}The longitudinal spin-spin correlation $C(r)$ are shown 
along the zigzag leg with the reference spin situated on same zigzag ladder in (a), whereas $C(r)$ on 
the other zigzag leg is shown in (b) for $J_2=0.1$ and five values of $J_3 = 0.1, 0.3, 
0.5, 0.7$ and $0.9$ with $N = 122$. In (c) and (d), $C(r)$ in the same zigzag leg  
are shown for $J_2 = 0.23$ and $0.3$ with same five values of $J_3$. The 
solid lines represent respective exponential fits.
}
\end{figure}

\subsection{\label{sec22}Spin-spin correlation C(r)}
We calculate the longitudinal spin-spin correlation $C(r)=<S^z_0 S^z_r>$,
where $S^z_0$ and $S^z_r$ are the $z$-component of the spin operators at the reference
site $0$ chosen at the middle of a zigzag chain and the site $r$ at a distance $r$ from $0$th spin, respectively.
In Fig.~\ref{fig:figure1}, the distance $r$ is shown along the same zigzag ladder with bold numerics
with respect to the reference site $0$, whereas, normal numerics
represent distances on the other zigzag leg. We note that in 
$J_2/|J_1|< \frac{1}{4}$ limit, all the spins are aligned parallelly
on individual zigzag ladder and have short range longitudinal correlation for finite $J_3$. $C(r)$ follows an exponential behavior
as shown in the Fig.~\ref{fig:cor}(a) for $J_2 = 0.1$, and $J_3 = 0.1,0.3,0.5,0.7$ and $0.9$. We notice that each zigzag ladder shows
collinear arrangement of spins as $C(r)>0$, but it decays exponentially with $r$ i.e.,

\begin{eqnarray}
\label{eq:exp1}
 C(r) \propto \exp \left (\frac{-r}{\xi} \right).
\end{eqnarray}

\begin{figure}
\centering
\includegraphics[width=3.0 in]{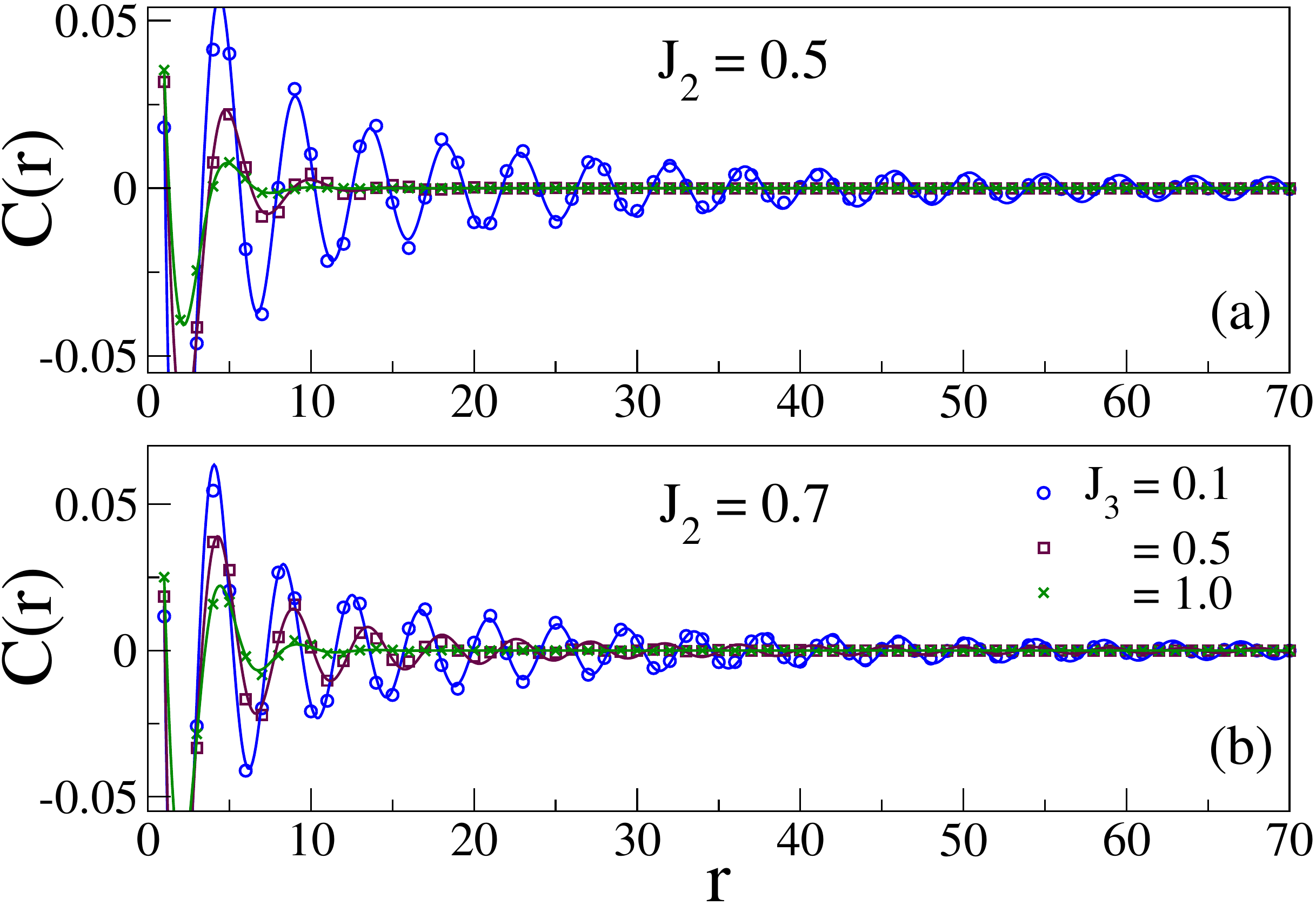}
\caption{\label{fig:cor2} For three values of $J_3=0.1, 0.5$ and $1.0$ with $N=298$, $C(r)$ are shown in (a) and 
(b) for $J_2=0.5$ and $0.7$, respectively. The solid curves represent respective sinusoidal fits with algebraic 
or exponential decay.
}
\end{figure}

The fitting curve represents an exponential 
function with correlation length $\xi$.
Fig.~\ref{fig:cor}(b) shows the $C(r)$ of the same reference spin with the spins on the other zigzag leg.
The negative values suggest anti-parallel arrangement of spins relative to the reference spin leg. This behavior of $C(r)$
confirms the stripe nature of spin arrangement on each zigzag ladder. Therefore, we call it striped phase.
On further increase in $J_2$, $C(r)$ starts to oscillate at higher $J_3$ even at the limit $J_2 < 0.25$.
For $J_2=0.23$, C(r) is shown in Fig.~\ref{fig:cor}(c) for the same set of $J_3$ values. We note that NC(SRO)
arises for $J_3 \ge 0.3$. While C(r) for $J_3=0.1$ is fitted by Eq.~\ref{eq:exp1}, C(r) for other $J_3$ can be fitted
with the equation below,

\begin{eqnarray}
\label{eq:exp2}
C(r) \propto \exp \left({\frac{-r}{\xi}}\right)  \sin \left(\theta r+c\right).
\end{eqnarray}

The NC order can be easily noticed at lower $J_3$ for $J_2 > 0.25$.
For $J_2=0.3$, C(r) is shown in Fig.~\ref{fig:cor}(d) and fitted by Eq.~\ref{eq:exp2}.
We note that $\xi$ decreases with $J_3$.
For moderate $J_2$, the NC phase follows SRO behavior, whereas it shows QLRO in the gs for higher $J_2>0.45$
but for small $J_3$. The transition between NC(SRO) to NC(QLRO) seems continuous, and hence it is difficult
to find an accurate phase boundary. In QLRO regime C(r) is fitted with sinusoidal power law function written as
\begin{eqnarray}
\label{eq:pow}
C(r) \propto r^{-\kappa} \sin \left(\theta r+c\right).
\end{eqnarray}

In Fig.~\ref{fig:cor2}(a) and (b), $C(r)$ are plotted for $J_2=0.5$ and $0.7$,
respectively, with $J_3=0.1,0.5$ and 1.0.
For $J_2=0.5$, and $J_3=0.1$, C(r) fits with power law in Eq.~\ref{eq:pow} where $\kappa \approx 1$, whereas
C(r) follows exponential decay at $J_3=0.5$ and 1.0 with $\xi = 2.29$ and 1.56, respectively. For $J_2=0.7$, and $J_3=0.1$ and 0.5, 
C(r) decays algebraically with $\kappa= 1.15$ and 1.37, respectively,
but exponentially for $J_3=1.0$ with $\xi = 1.99$. We notice that the width of the NC(QLRO)
region increases with $J_2$.

\begin{figure}
\centering
\includegraphics[width=3.0 in]{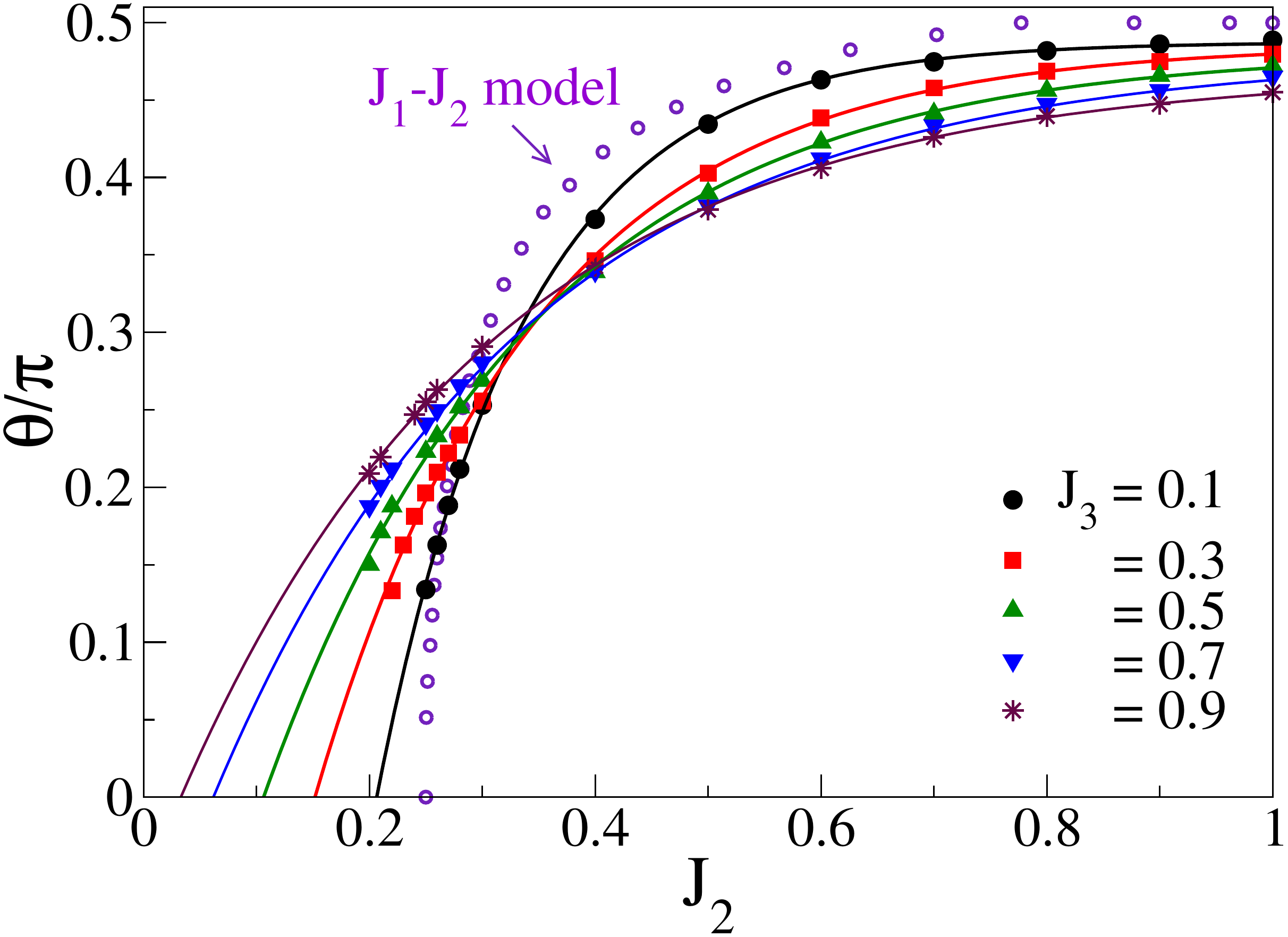}
\caption{\label{fig:pitch} The variation of pitch angle $\theta$ with $J_2$ are shown 
for five values of $J_3 = 0.1,0.3,0.5,0.7$ and $0.9$. The open circles represent $\theta$ for 
$J_1-J_2$ spin-1/2 model on a zigzag ladder with FM $J_1$ and AFM $J_2$.
}
\end{figure}

\subsection{\label{sec23}Pitch angle $\theta$}

In the NC phase we calculate pitch angle $\theta$ from the fitting parameter in Eq.~\ref{eq:exp2} and Eq.~\ref{eq:pow}.
$\theta/\pi$ is plotted as a function of $J_2$ for various values of $J_3$, as shown in Fig.~\ref{fig:pitch}. $\theta/\pi$ versus $J_2$
curves are fitted with function $\theta/\pi=a[1-e^{{b(J_2-J^c_2)}}]$,
where $a$, $b$ and $J^c_2$ are the fitting parameters. $J^c_2$ is the phase boundary point between CS(SRO) and NC(SRO) phases for a given
$J_2$ and $J_3$. $\theta$ increases from 0 to $\pi/2$ with $J_2$. The $\theta$ $\approx$ $\pi/2$
region is confined to high $J_2/J_3$ limit. The variation of $\theta$ is represented by color gradient in the phase diagram
in Fig.~\ref{fig:phase}.

\subsection{\label{sec24}Correlation length $\xi$}
\begin{figure}
\centering
\includegraphics[width=3.0 in]{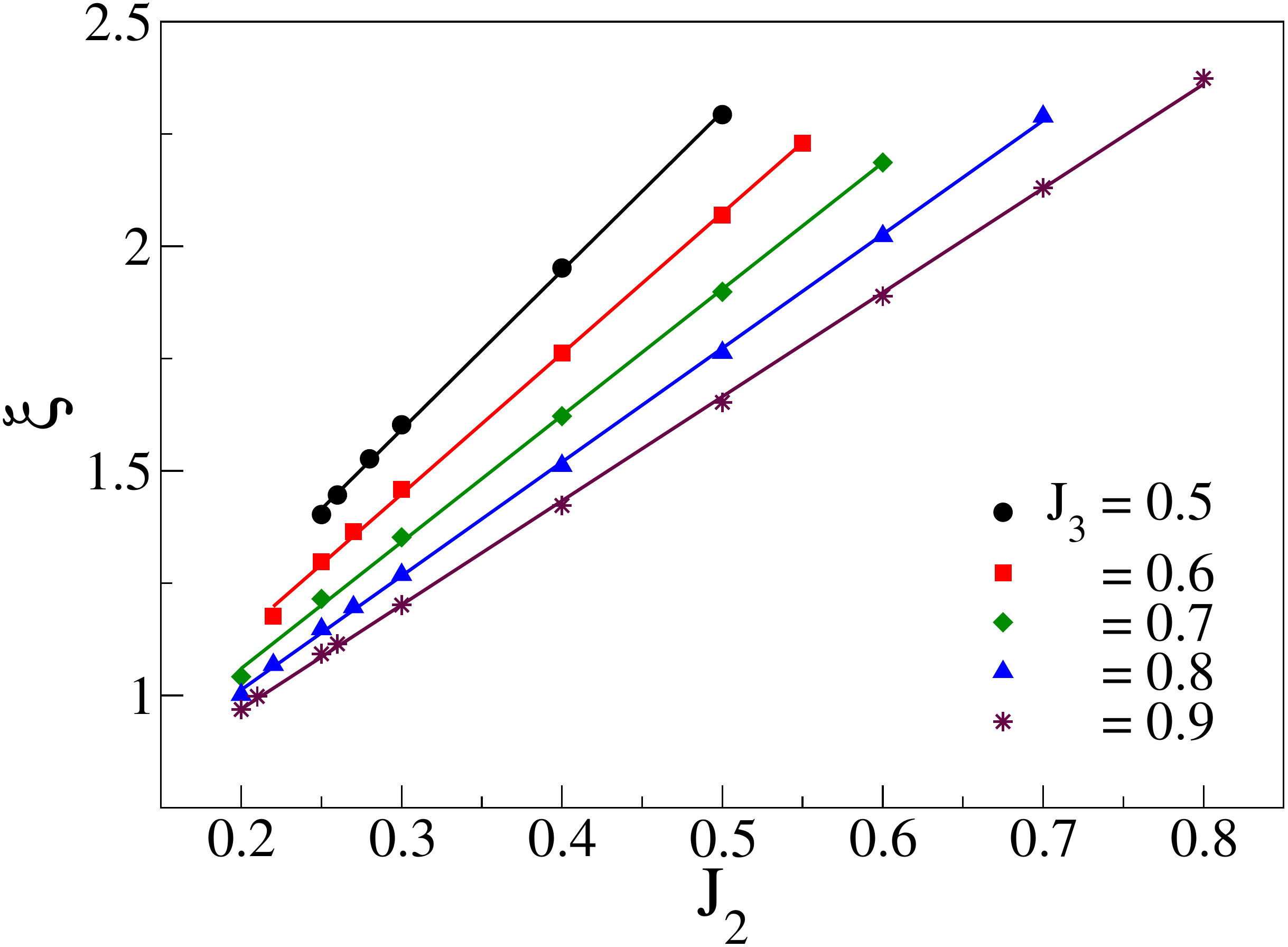}
\caption{\label{fig:cor_leng} In NC(SRO) phase $\xi-J_2$ curves are
shown for five values of $J_3 = 0.5, 0.6, 0.7, 0.8$ and $0.9$. The solid lines represent respective linear fits.
}
\end{figure}

The correlation length $\xi$ extracted from fitting Eq.~\ref{eq:exp1} and Eq.~\ref{eq:exp2} is a measure of
correlation length in CS(SRO) and NC(SRO) phase, respectively. The nature of $\xi$ in CS(SRO) phase is discussed in ref.~\cite{MAITI2019}. 
In NC(SRO) regime $\xi$ are plotted as function of $J_2$ for $J_3 = 0.5,0.6,0.7,0.8,0.9$ in Fig.~\ref{fig:cor_leng}.
The correlation length can be fitted by $\xi=c+dJ_2$, where $c$ and $d$ are the fitting parameters. We note that $\xi$ increases
with $J_2/J_3$. Higher value of $J_2$ needs more strength in $J_3$ to keep the same correlation length in NC(SRO) phase. 
Surprisingly, this behavior is completely opposite in the case of CS(SRO) phase, where higher $J_2$ requires lower $J_3$ to
sustain the same correlation length ~\cite{MAITI2019}. When $\xi \le 1$, dominant correlation strengths
become confined within the three nearest neighbors among which the rung bond correlation is dominant over other two bond strengths.
In fact $\xi \le 1$ represents the correlation length within nearest neighbor distance; as per our convention
of distance, both $r=1$ and $r=2$ are the nearest neighbors
to the reference spin.
In this limit, the system behaves like a collection of singlet rung dimers. The 
varying strength of nearest neighbor bond correlations
depending on $J_2$ and $J_3$ are discussed in the next subsection.

\begin{figure}[b]
\centering
\includegraphics[width=3.0 in]{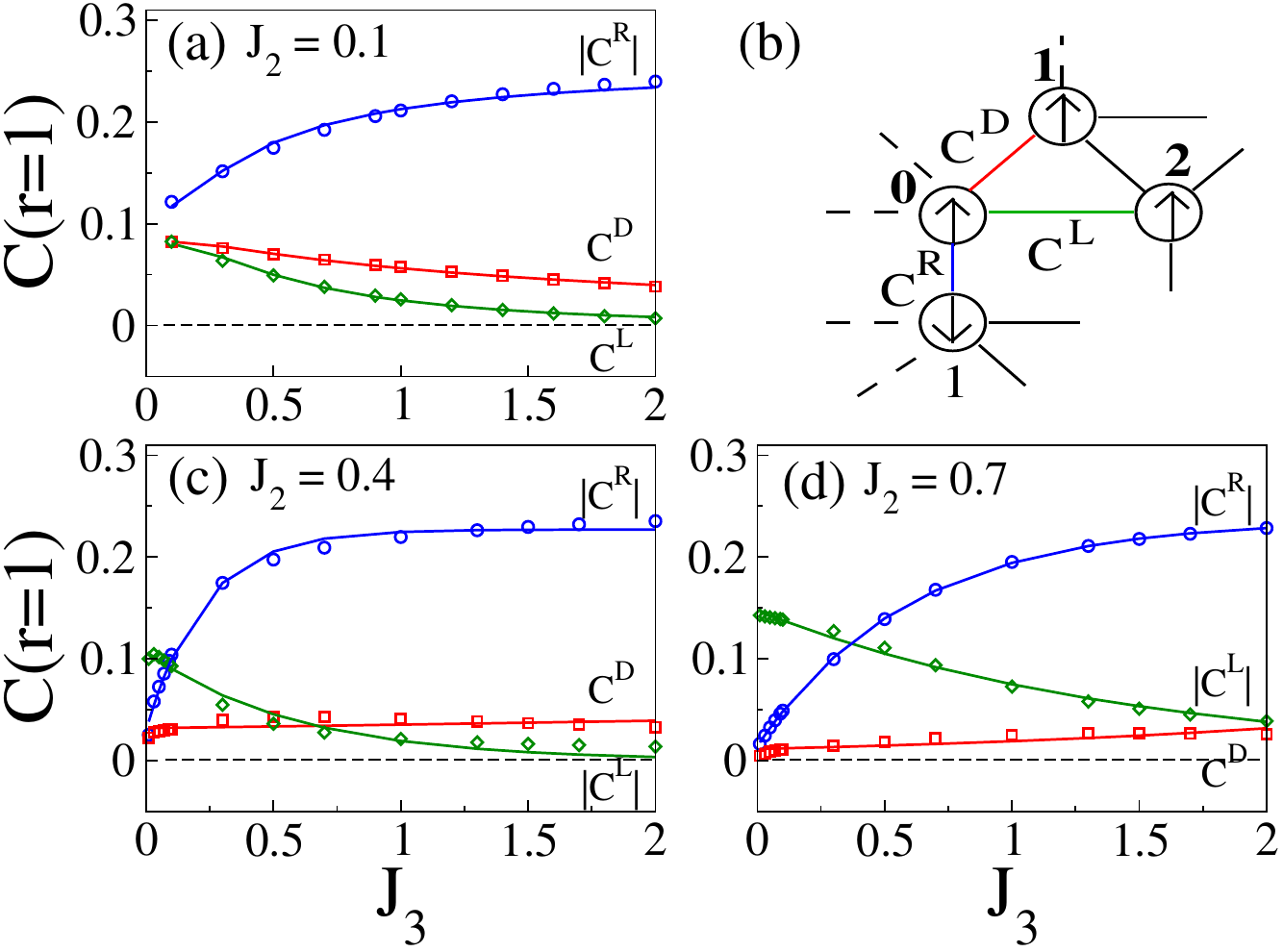}
\caption{\label{fig:bond}
Nearest neighbor correlation function $C(r=1)$ at the mid of zigzag ladder is shown. 
The $C(r=1)$ along the rung ($C^R$, circle), diagonal direction ($C^D$, square), and 
leg ($C^L$, diamond) are shown in the schematic in (b).  $C(r=1)-J_3$ plots 
are shown for $J_2 = 0.1, 0.4$ and $0.7$ in (a), (c) and (d), respectively. 
The lines represent respective exponential fits.
}
\end{figure}

\subsection{\label{sec25}Nearest neighbor bond correlation $C(r=1)$}
It is quite interesting to see the relative strength of nearest neighbor $C(r=1)$
or longitudinal bond order in the parameter space. The magnitude of $C(r=1)$ along the rung $|C^R|$, along the leg  $|C^L|$, and along the
zigzag leg $C^D$ are shown for $J_2/|J_1| = 0.1, 0.4$, and $0.7$ in Fig.~\ref{fig:bond}(a), (c) and (d), respectively.
The bonds along three directions are shown in the schematic Fig.~\ref{fig:bond}(b). We notice that $C^D$
and $C^L$ have positive values for CS(SRO) phase whereas, $C^L$ becomes negative for NC phase.
In the NC phase $|C^L|$ is dominant for small $J_3$, but $|C^R|$ dominates for $J_3 > 0.08$ and $0.38$
for $J_2$ = 0.4 and 0.7,respectively. The effect of $J_3$ on $C^D$ is weak, and also the magnitude of $C^D$ is small.
Therefore, we can safely conclude that major contributions of energy come from $|C^R|$ and $|C^L|$ in NC phase.
$|C^R|$ increases exponentially with $J_3$ and saturates to a value which is nearly equal to 0.25.

\section{\label{sec4}Linear spin wave analysis}
In the CS(SRO) phase spins on the same zigzag ladder are arranged
ferromagnetically, whereas spins from different zigzag ladders are arranged antiferromagnetically to each other.
We perform the linear spin wave analysis of the Hamiltonian for this phase. 
We use the Holstein-Primakoff transformation to the Hamiltonian in Eq.~\ref{eq:ham}.
The details of the calculation are given in appendix A.

The Hamiltonian can be written in terms of bosonic operators $a_{j}$,$b_{j}$,$a^{+}_{j}$ and $b^{+}_{j}$, where $a_{j}$/$a^{+}_{j}$ and
$b_{j}$/$b^{+}_{j}$ correspond to spin up and spin down operators or spins on leg $l=1$ and $l=2$, respectively.
We consider only up to quadratic terms. After Fourier transformation, the resultant Hamiltonian can be written as

\begin{eqnarray}
\label{eq:ham1}
H = (2J_{1}+2J_{2}-J_{3})Ns^{2} + \sum_{k} s [(2J_{1}(\cos{k}-1) \nonumber \\
+2J_{2}(\cos{2k}-1)+J_{3})(a^{+}_{k}a_{k}+b^{+}_{k}b_{k}) \nonumber \\
+J_3(a^{+}_{k}b^{+}_{-k}+a_{k}b_{-k})]. 
\end{eqnarray}

The above Hamiltonian can be transformed to diagonal form using the Bogoliubov transformation i.e.,
\begin{eqnarray}
\label{eq:bog}
a_{k} = uc_{k}-vd^{+}_{k}, \nonumber \\
b^{+}_{-k} = -vc_{k}+ud^{+}_{k},
\end{eqnarray}

where $u^{2}-v^{2}=1$, $u^{2}+v^{2}=\frac{J_{k}}{\sqrt{J^{2}_{k}-J^{2}_{3}}}$ and $2uv=\frac{J_{3}}{\sqrt{J^{2}_{k}-J^{2}_{3}}}$, and 
$J_{k}=2J_{1}(\cos{k}-1)+2J_{2}(\cos{2k}-1)+J_{3}$.
Applying Bogoliubov transformation, we get

\begin{eqnarray}
H = (2J_{1}+2J_{2}-J_{3})Ns^{2} \nonumber \\
+ \sum_{k} \omega_{k}(c^{+}_{k}c_{k}+d^{+}_{k}d_{k}+1), 
\end{eqnarray}
where $\omega_{k}= S(\sqrt{J^{2}_{k}-J^{2}_{3}})$.

The gs energy per bond is given by
\begin{eqnarray}
\epsilon = (J_{1}+J_{2}-\frac{J_{3}}{2})S(S+1) \nonumber \\ 
+ \sum_{k} \frac{s}{2\pi}\int_{0}^{\pi} \sqrt{J^{2}_{k}-J^{2}_{3}} dk
\end{eqnarray}

The $\epsilon$ can be minimized using $\frac{d\omega_{k}}{dk}=0$ and we find
these conditions; $\cos{k}=\frac{-J_{1}}{4J_{2}}$ and $\cos{k}=\frac{-J_{1}}{4J_{2}}\pm\frac{\sqrt{(J_{1}+4J_{2})^2-4J_{2}J_{3}}}{4J_{2}}$.
The second condition $ J_{3} \le \frac{(J_{1}+4J_{2})^2}{4J_2}$ for any real value of $\cos{k}$, gives the phase boundary between
CS(SRO) and NC(SRO) phases. This boundary is similar to that found by DMRG calculation.

\begin{figure}
\centering
\includegraphics[width=3.0 in]{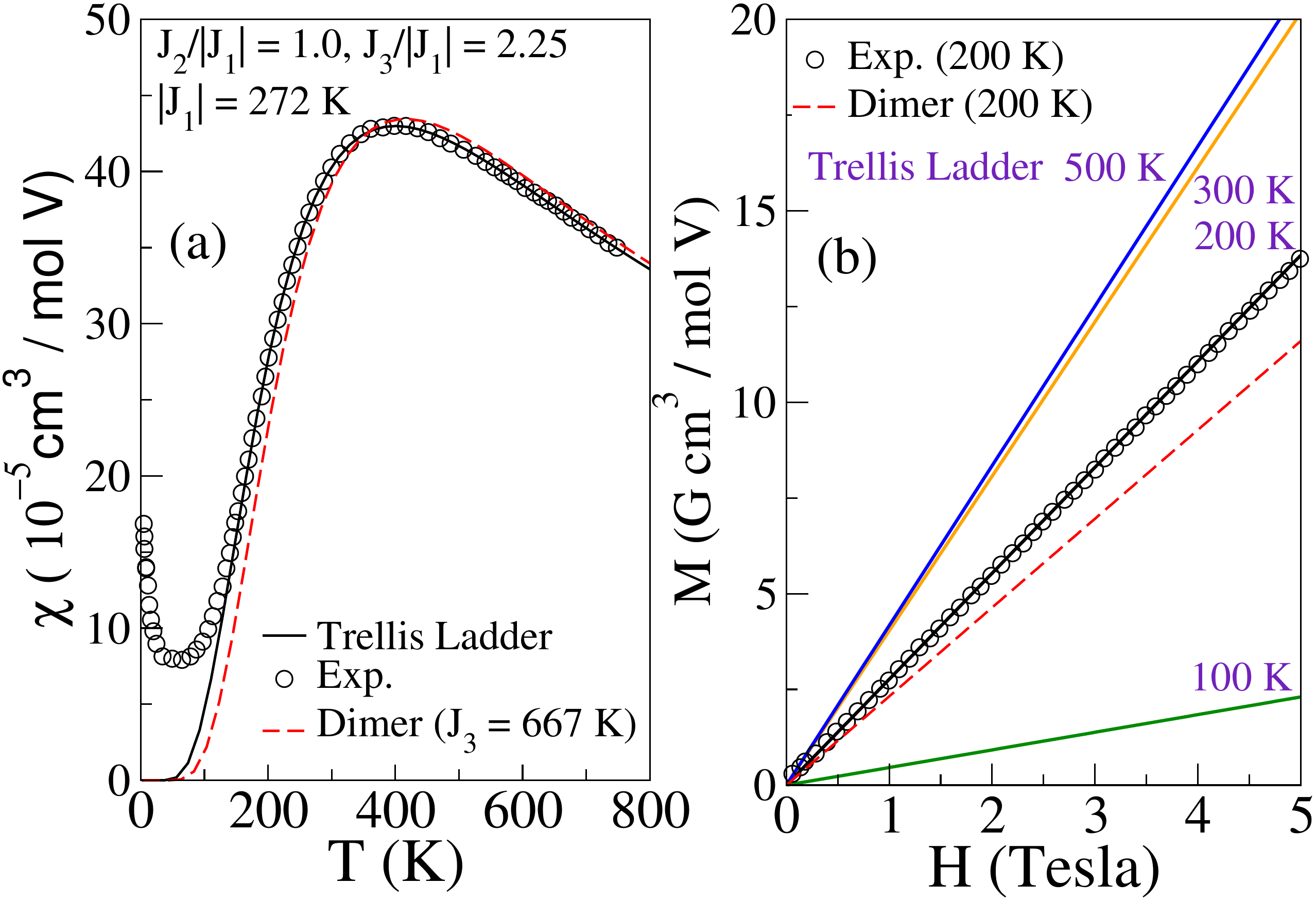}
\caption{\label{fig:sus_mag}(a) Magnetic susceptibility $\chi$ as a function of temperature $T$ for 
CaV$_2$O$_5$ sample 1~\cite{Johnston2000} is shown by the circles. Solid curve represents the fitted curve obtained 
by the trellis ladder model and dashed curve represents the fitted curve using dimer model.
(b) Circles represent Magnetization $M$ versus applied magnetic field $H$ 
curve at $T$ = 200 K for CaV$_2$O$_5$ sample 2~\cite{Johnston2000}. The black solid line is the fit using
our model and dashed line represents the fit for a perfect dimer system at $T=200K$. The fitting parameters are same
as used to fit $\chi-T$ curve. The other $M-H$ plots 
for $T = 100 K, 300 K$ and $500 K$ are shown by the solid lines using the model in Eq.~\ref{eq:ham}.
}
\end{figure}

\begin{figure}[b]
\centering
\includegraphics[width=3.0 in]{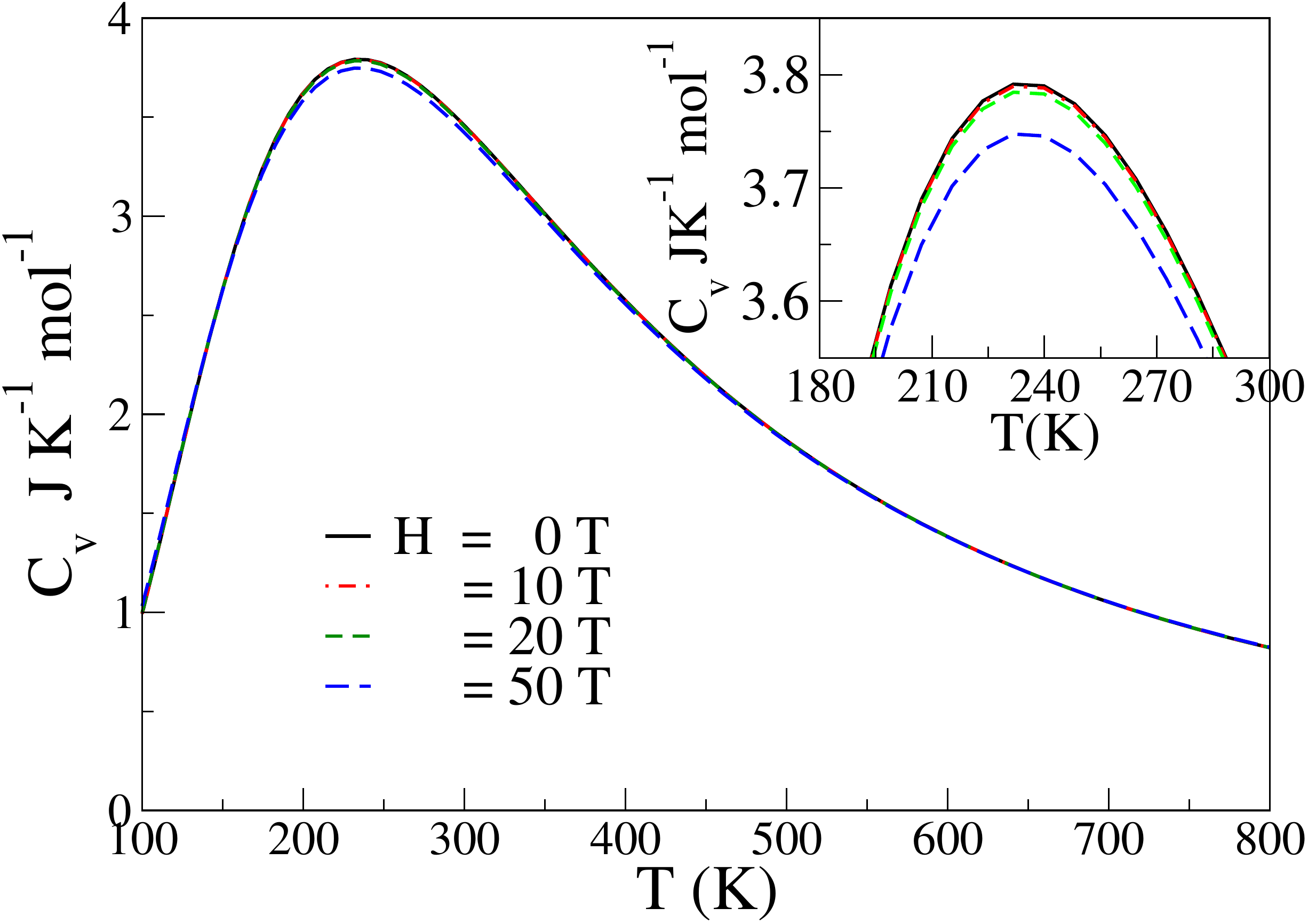}
\caption{\label{fig:cv}Specific heat $C_v(T)$ are plotted as function of $T$ with $J_2/|J_1|=1.0$, $J_3/|J_1|=2.25$ and
$|J_1|=272 K$ for $H$ = 0, 10, 20 and 50 Tesla. 
The zoomed $C_v(T)$ near the peak are shown in the inset.
}
\end{figure}

\section{\label{sec5}Fitting experimental data of CaV$_2$O$_5$}
There are many vanadate compounds like CaV$_2$O$_5$, MgV$_2$O$_5$, 
NaV$_2$O$_5$ etc., which are suspected to behave effectively like two leg ladders coupled by 
zigzag bonds forming trellis lattice like structure. Among these 
materials the interladder coupling ($J_1$) in CaV$_2$O$_5$ is expected to be 
ferromagnetic. The LDA+U calculations performed by Korotin {\it et al.}~\cite{Tdasgupta2000} 
give an estimation of the $J_1$ ,$J_2$, and $J_3$ exchange interaction strengths as $-28 K$, 
$122 K$ and $608 K$, respectively. In this compound, $V^{4+}$ ions have one electron in 
d-orbital and behave like spin-1/2 ions. The experimental magnetic susceptibility 
$\chi(T)$ is taken from sample 1 and magnetization $M(H)$ is taken from sample 2 of 
ref.~\cite{Johnston2000} which are represented by circles in Fig.~\ref{fig:sus_mag}(a) and (b), respectively. 
The dimer model fitting of susceptibility data deviates significantly from the experimentally observed data. 
The experimental data is shown as circle and dimer fit is shown by red dashed line in Fig.~\ref{fig:sus_mag}(a). 
The model Hamiltonian in Eq.~\ref{eq:ham} 
is used with $J_1 = -272 K$, $J_2 = 272 K$ and $J_3 = 612 K$ to fit the experimental data of $M(H)$ 
and $\chi(T)$.  The fitting curve of $\chi(T)$ shown by black solid curve is in excellent agreement with
experimental data for $T > 160 K$. As shown in Fig.~\ref{fig:sus_mag}(b) $M-H$ curve fitted with dimer 
model shown by red dashed line is quite off at high $H$, whereas our model gives excellent 
fitting, as shown by the black solid line at $T = 200 K$. We predict $M-H$ curve at other three different 
$T= 100, 300$ and $500 K$. We notice the enhancement of $M$ as a function of $T$, which is
 quite unusual. This behavior of $M-H$ curve can be understood in terms of
large singlet-triplet gap. A moderate temperature enhances the possibility to reach 
higher magnetic state for a given field $H$.

We also predict the magnitude of specific heat $C_v$ as a function of $T$ for four values of 
magnetic field $H = 0, 10, 20, 50$ Tesla as shown in  Fig.\ref{fig:cv}. The $C_v$ has broad peak 
at $T \approx 235 K$. The effect of magnetic field $H$ is small. The $C_v$ decreases with $H$,
but the suppression of $C_v$ is visible only near the peak. Initially CaV$_2$O$_5$ was assumed 
to be only a dimer system with singlet-triplet energy gap ~ $660 K$ ~\cite{ONODA1996}.
We use the model Hamiltonian in Eq.~\ref{eq:ham}, and our fittings of $\chi(T)$ and 
$M(H)$ with same model parameters suggest that $J_1$ and $J_2$ are only 1/2 of $J_3$. 
It is found that our predicted values of $J_1$ and $J_2$ are significantly different from 
the predicted values in ref.~\cite{Tdasgupta2000}, whereas the value of $J_3$ is similar 
with their calculated value by LDA+U method. 

\section{\label{sec6}Discussion and Conclusions}
In this paper we have studied the isotropic Heisenberg spin-1/2 model, 
given in Eq.~\ref{eq:ham}, on the trellis ladder. The QPD
of this model is constructed. The phase boundaries of the QPD
are calculated based on the correlation function $C(r)$, pitch angle $\theta$
and correlation length $\xi$ using the DMRG method. Our linear spin
wave analysis of this model predicts phase boundary of CS(SRO) and NC(SRO) phases,
and it is quite consistent with our DMRG results.
We also use this model to fit $\chi-T$ and $M-H$ data of CaV$_2$O$_5$, and understand the
temperature $T$ dependence of $M-H$ curves and magnetic field $H$ dependence of $C_v-T$ curves.

In fact our lattice system can also be mapped to a two 
coupled $J_1-J_2$ Heisenberg spin-1/2 chains. Zinke {\it et al.} studied the effect 
of interchain coupling $J_3$ on non-collinear phase in a coupled 2D array
of $J_1-J_2$ spin chains using the coupled cluster theory ~\cite{Zinke2009}.
They showed that the collinear to non-collinear transition point $J^c_2$
increases with $J_3$. However, our model shows that the critical value $J^c_2$ decreases with $J_3$. This
inconsistency may be because of the confined geometry or ladder structure in our case. 
The $J^c_2$ value at phase boundary of CS(SRO) and NC(SRO) phases
decreases with $J_3$, and it can also be shown by linear spin wave analysis. 
As shown in Fig.~\ref{fig:pitch}, the variation of $\theta$ for $J_2 > 0.3$ decreases with $J_3$ and this 
trend is consistent with literature ~\cite{Zinke2009}, and this may happen because of
the deconfinement of quasi-particle along rung of the model. In Fig.~\ref{fig:phase} of QPD
the majority of the parameter space is SRO phase which is basically gapped spin liquid phase~\cite{Mila2000,dagotto96}; however,
for small value of $J_3/J_2$, an incommensurate (QLRO) phase appears, which is quite unique in
this ladder system. The $J_1-J_2$ spin-1/2 zigzag model in similar parameter space
shows either incommensurate (SRO) or decoupled phase ~\cite{mk2015,white96}.
The QLRO in the system may be induced because of dominant effective anti-ferromagnetic
interaction along the leg.

We apply this model to understand the magnetic properties of the CaV$_2$O$_5$, and have 
reliable fitting of the experimental data~\cite{Johnston2000}. 
We apply a criterion of simultaneous fitting of both experimental $\chi-T$ and $M-H$ curves.
Our best fit suggests that $J_2/|J_1| $ is close to 1, and $J_1$ is approximately $-272 K$.
For a given $H$, $M(H)$ for this system increases with $T$, whereas in general magnetization 
decreases with increasing temperature. We notice that in a highly gapped system, 
higher $T$ allows the system to access the higher magnetic states easily;
therefore, it is much easier to magnetize this system at moderate temperature for a given $H$. 
Our calculated singlet-triplet energy gap is  $~459 K$, whereas dimer model predicts it as ~ $660 K$.
The Knight shift and spin-lattice relaxation measurements for CaV$_2$O$_5$ show 
energy gaps are $464 K$ and $616 K$, respectively ~\cite{Iwase1996}. Our predicted energy gap is closer to the Knight shift measurement.     
The modelling of $\chi(T)$ of CaV$_2$O$_5$ was done by Miyahara {\it et al.}
using QMC method, and they showed that small $J_1$ does not effect the magnetic $\chi(T)$,
as shown in Fig.6 of ref.~\cite{Miyahara1998}. They estimated the value of $J_1=45K$, $J_2=67 K$
and $J_3=672 K$. Johnston {\it et al.} treated this system as collection of dimers, and extracted the value of
$J_3=667 K$ with small $J_1$ and $J_2$ ~\cite{Johnston2000}. Korotin {\it et al.} also calculated the value
of $J_1=-28 K$, $J_2=122 K$ and $J_3=608 K$; however, their calculation 
also assumes other types of interactions ~\cite{Tdasgupta2000}.
Our simultaneous fitting of experimental $\chi-T$ and $M-H$
data also suggests it as dominant dimer with $J_3=612 K$, but -$J_1$ and $J_2$ are only about half in magnitude of the $J_3$.

In summary, we study the QPD of model Hamiltonian in Eq.~\ref{eq:ham} on the trellis ladder. We show that
$J_3$ plays an important role to localize the system. This system shows interesting CS(SRO) and NC(QLRO) which 
is rare in ladder like structures. This model Hamiltonian 
is used to fit the experimental magnetic properties of CaV$_2$O$_5$ and we also show that the interaction $J_1$ and $J_2$ are much larger
than earlier predicted values, and $J_1$ is ferromagnetic in nature.
In many zigzag ladder systems like LiCuVO$_4$~\cite{Mourigal2012}, Li$_2$CuZrO$_4$~\cite{Drechsler2007},
Li$_2$CuSbO$_4$ ~\cite{dutton_prl} etc.,
where three dimensional ordering occurs at low $T$,  this model 
can be applied to understand the effect of interladder
coupling in the system. We have also predicted the $M-H$ and $C_v-T$ curves which can be verified experimentally.

\textbf{Acknowledgements} MK thanks Prof. Z. G. Soos for useful discussions and also thanks DST 
for a Ramanujan Fellowship SR/S2/RJN-69/2012
and DST India for funding computation facility through SNB/MK/14-15/137. Debasmita thanks Dayasindhu Dey for help 
in linearised spin wave theory  calculations.  

\section{\label{sec7}Appendix A}
For up spins the Holstein-Primakoff transformations take the form

\begin{eqnarray}
S^{z}_{Aj} = s-a^{+}_{j}a_{j}, \nonumber \\
S^{+}_{Aj} = \sqrt{(2s-a^{+}_{j}a_{j})}a_{j}, \nonumber \\
S^{-}_{Aj} = a^{+}_{j}\sqrt{(2s-a^{+}_{j}a_{j})} ,
\end{eqnarray}

For the down spin

\begin{eqnarray}
S^{z}_{Bj} = -s+b^{+}_{j}b_{j}, \nonumber \\
S^{+}_{Bj} = a^{+}_{j}\sqrt{(2s-a^{+}_{j}a_{j})}, \nonumber \\
S^{-}_{Bj} = \sqrt{(2s-a^{+}_{j}a_{j})}a_{j} ,
\end{eqnarray}

We use the linear approximation at classical limit

\begin{eqnarray}
S^{z}_{Aj} = s-a^{+}_{j}a_{j}, \nonumber \\
S^{+}_{Aj} = \sqrt{2s} a_{j}, \nonumber \\
S^{-}_{Aj} = \sqrt{2s} a^{+}_{j},
\end{eqnarray}

for spin up, and for spin down

\begin{eqnarray}
S^{z}_{Bj} = s-b^{+}_{j}b_{j}, \nonumber \\
S^{+}_{Bj} = \sqrt{2s} b^{+}_{j}, \nonumber \\
S^{-}_{Bj} = \sqrt{2s} b_{j}.
\end{eqnarray}

In terms of bosonic operators, the Hamiltonian takes the form upto quadratic order as

\begin{eqnarray}
\label{eq:ham3}
H = (2J_{1}+2J_{2}-J_{3})Ns^{2} + \sum_{j} s [[J_{1}(a^{+}_{j}a_{j+1}+b^{+}_{j}b_{j+1}) \nonumber \\ 
+J_{2}(a^{+}_{j}a_{j+2}+b^{+}_{j}b_{j+2})+J_{3} a_{j}b_{j} + h.c.] \nonumber \\
-(J_{1}+ J_{2})(a^{+}_{j}a_{j}+b^{+}_{j}b_{j})-J_{1}(a^{+}_{j+1}a_{j+1}+b^{+}_{j+1}b_{j+1}) \nonumber \\
+J_{2}(a^{+}_{j+2}a_{j+2}+b^{+}_{j+2}b_{j+2})+J_3(a^{+}_{j}a_{j}+b^{+}_{j}b_{j})]. 
\end{eqnarray}

Fourier transforms of the bosonic operators are,

\begin{eqnarray}
a_{j} = \sum_{k} \exp{(-ikj)}  a_{k}, \nonumber \\
a^{+}_{j} = \sum_{k} \exp{(ikj)}  a^{+}_{k}.
\end{eqnarray}


\bibliography{references}

\end{document}